\documentclass[prl,twocolumn]{revtex4}
\usepackage{epsfig}
\usepackage{amsmath}
\usepackage{epsfig}

\begin{document}
\title
{Pearson's correlation coefficient in the theory of networks: A comment}    
\author{Zafar Ahmed$^1$ and Sachin Kumar$^2$}
\affiliation{$^1$Nuclear Physics Division, $^2$Theoretical Physics Section,  Bhabha Atomic Research Centre, Mumbai 400085, India}
\email{1:zahmed@barc.gov.in, 2:sachinv@barc.gov.in}
\date{\today}
\begin{abstract}
In statistics, the  Pearson correlation coefficient $r_{x,y}$ determines the degree of linear correlation between two variables and it is known that $-1 \le r_{x,y} \le 1$. In the theory of networks, a curious expression proposed in [PRL {\bf 89} 208701 (2002)] for degree-degree correlation coefficient $r_{j_i,k_i}, i\in [1,M]$ has been in use. We realize that the suggested form is the conventional Pearson's coefficient for $\{(j_i,k_i), (k_i,j_i)\}$ for $2M$ data points and hence it is rightly dedicated to undirected networks.
\end{abstract}
\maketitle
In statistics [1],  the Pearson correlation coefficient $r_{x,y}$ determines the degree of linear correlation between two variables $x$ and $y$, 
given the data $x_i$ and $y_i$, $i\in [1,n]$. The correlation coefficient is defined as $r_{x,y}=\frac{\mbox{Cov}(x,y)}{\sigma_x \sigma_y}$,
\begin{eqnarray}
\mbox{Cov}(x,y){=}\frac{1}{n}\sum_{i=1}^{n} x_iy_i -\bar x \bar y, \quad \sigma^2_x{=}\frac{1}{n} \sum_{i=1}^{n} x^2_i{-} (\bar x)^2,
\end{eqnarray}
where $\mbox{Cov}(x,y)$ is called the co-variance of $x_i$ and $y_i$, $\sigma_x$ is the standard deviation of $x_i$ and  $\bar x$ is the arithmetic mean $\bar x = \frac{1}{n}\sum_{i=1}^{n} x_i$. The correlation coefficient is usually written [1] as
\begin{small}
\begin{equation}
r_{x,y}=\frac{n^{-1}\sum_{i=1}^n x_i y_i -\bar x \bar y}{\sqrt{n^{-1}\sum_{i=1}^{n} x^2_i - (\bar x)^2}\sqrt{n^{-1}\sum_{i=1}^{n} y^2_i -(\bar y)^2}}.
\end{equation}
\end{small}
If the data points follow $y_i=\pm m x_i+c$  $\forall$  $i$ $\in (1,n)$ for fixed values of $m>0$ and $c$, $r_{x,y}=\pm 1$, otherwise we have $-1<r_{x,y} <1.$

In the theory of networks [2],
let $j_i$ and $k_i$ be  the  excess  in-degree  and  out-degree  of
the  vertices  that  the
$i^{th}$  edge  leads  into  and  out  of  respectively, and
$M$ is the number of edges. The degree-degree correlation coefficient can be defined conventionally (2) as
\begin{small}
\begin{eqnarray}
r_{j,k}&=&\frac{M^{-1}}{\sigma_j \sigma_k} \left(\sum_{i=1}^M j_i k_i -\bar j \bar k \right) \nonumber \\
r_{j,k}&=&\frac{M^{-1}\sum_{i=1}^M j_i k_i -\bar j \bar k}{\sqrt{M^{-1}\sum_{i=1}^{M} j^2_i -(\bar  j)^2} \sqrt{M^{-1}\sum_{i=1}^{M} k^2_i -(\bar k)^2}}.
\end{eqnarray}
\end{small}
curiously, in Ref. [3], $r_{j,k}$ has been  proposed  as 
\begin{equation}
r_{j,k}{=}\frac{M^{-1}\sum_{i=1}^M j_i k_i{-} [M^{-1} \frac{1}{2} \sum_{i=1}^M (j_i{+}k_i)]^2}{M^{-1}\frac{1}{2}\sum_{i=1}^{M} (j^2_i{+}k^2_i) {-}[ M^{-1}\frac{1}{2}\sum_{i=1}^{M} (j_i{+}k_i)]^2}.
\end{equation}
This can be re-written to look much close to Eq. (3) as
%\begin{small}
\begin{equation}
r_{j,k}=\frac{M^{-1}\sum_{i=1}^M j_i k_i -\bar j \bar k/2-((\bar j)^2+(\bar k)^2)/4}{(\sigma^2_j+\sigma^2_k)/2+(\bar j-\bar k)^2/4}.
\end{equation}
%\end{small}
In the trivial case of the perfect correlation when $j_i=k_i$,  all three Eqs. (3-5)  give $r_{j,k}=1$, incidentally. However, for the other case of the perfect linear correlation when $k_i=2j_i+1$, for $M=9$ points, we find that Eq. (3) gives 1 correctly, whereas Eqs.(4,5) give $r_{j,k}=13/77$. Next, when there is a quadratic dependence such as $k_i=j^2_i$, the  Eq. (3) gives $r_{j,k}=\sqrt{1500/1577}$ but Eqs. (4,5) give $r_{j,k}=-125/598$, a negative value. 

In another paper, the Eq. (4) has been used slightly mistakingly [5] as 
\begin{small}
\begin{equation}
r_{j,k}{=}\frac{M^{-1}\sum_{i=1}^M j_i k_i {-} M^{-1}  \sum_{i=1}^M \frac{1}{2}(j_i+k_i)^2}{M^{-1}\sum_{i=1}^{M}\frac{1}{2} (j^2_i{+}k^2_i) {-} M^{-1}\sum_{i=1}^{M} \frac{1}{2} (j_i{+}k_i)^2}.
\end{equation}
\end{small}
The Eq.(6) can be easily  reduced as
\begin{equation}
r_{j,k}=\frac{\sum_{i=1}^M (j^2_i+k^2_i)}{2\sum_{i=1}^M j_i k_i } = \frac{\sum_{i=1}^M (j_i -k_i)^2}{2\sum_{i=1}^M j_i k_i} +1 \ge 1,
\end{equation}

\noindent as $j_i,k_i>0$. The coefficient $r_{j,k}$ exceeds 1 and hence Eq. (6) fails to represent the correlation coefficient in any case.  Though in Eq. (26) of Ref. [4] a formula which is the same as the form (3) has been proposed, yet the use of Eq. (4) [3]  has been re-emphasized [4] for undirected networks. 

This apparent anomaly can be resolved by realizing that the interesting forms (4) and (5) are actually the conventional Pearson's coefficient (2) for the combined $2M$ data points in the case of undirected networks which actually are $\{(j_i,k_i), (k_i,j_i)\}, i \in [1,M]$. Thus, Eq. (3) is for directed and Eqs. (4,5) are for  un-directed networks. However, Eq. (6) [5] is a mistaken form of Eq. (4) or (5).
\vspace{-.5cm}
\section*{Acknowledgement:} We thank Dr. Sang Hoon Lee for a discussion.


\vspace{-.8cm}
\section*{References}
\vspace{-1.cm}
\begin{thebibliography}{5}
\bibitem{[1]} e.g., J. F. Kenny, Mathematics of Statistics (Part one) (D. Van Nostrand Co., Inc., New York, 1947) Chapter VIII.
\bibitem{[2]} M. E. J. Newman, Networks: An introduction (Oxford University Press, New York, 2010)	
\bibitem{[3]} M. E. J. Newman, Phys. Rev. Lett. {\bf 89}, 208701 (2002);
\bibitem{[4]} M. E. J. Newman, Phys. Rev. E {\bf 67}, 026126 (2003).
\bibitem{[5]} S. Jalan, A. Kumar, A. Zaikin and J. Kruth, Phys. Rev. E {\bf 94} 062202 (2016).
\end{thebibliography}
\end{document}